\documentstyle[twoside]{article}

\catcode`\@=11
\long\def\@makefntext#1{
\protect\noindent \hbox to 3.2pt {\hskip-.9pt  
$^{{\eightrm\@thefnmark}}$\hfil}#1\hfill}		

\def\@makefnmark{\hbox to 0pt{$^{\@thefnmark}$\hss}}	
	
\def\ps@myheadings{\let\@mkboth\@gobbletwo
\def\@oddhead{\hbox{}
\rightmark\hfil\eightrm\thepage}   
\def\@oddfoot{}\def\@evenhead{\eightrm\thepage\hfil
\leftmark\hbox{}}\def\@evenfoot{}
\def\sectionmark##1{}\def\subsectionmark##1{}}

\oddsidemargin=\evensidemargin
\newcounter{sectionc}\newcounter{subsectionc}\newcounter{subsubsectionc}
\renewcommand{\section}[1] {\vspace{12pt}\addtocounter{sectionc}{1} 
\setcounter{subsectionc}{0}\setcounter{subsubsectionc}{0}\noindent 
	{\tenbf\thesectionc. #1}\par\vspace{5pt}}
\renewcommand{\subsection}[1] {\vspace{12pt}\addtocounter{subsectionc}{1} 
	\setcounter{subsubsectionc}{0}\noindent 
	{\bf\thesectionc.\thesubsectionc. {\kern1pt \bfit #1}}\par\vspace{5pt}}
\renewcommand{\subsubsection}[1] {\vspace{12pt}\addtocounter{subsubsectionc}{1}
	\noindent{\tenrm\thesectionc.\thesubsectionc.\thesubsubsectionc.
	{\kern1pt \tenit #1}}\par\vspace{5pt}}
\newcommand{\nonumsection}[1] {\vspace{12pt}\noindent{\tenbf #1}
	\par\vspace{5pt}}

\newcounter{appendixc}
\newcounter{subappendixc}[appendixc]
\newcounter{subsubappendixc}[subappendixc]
\renewcommand{\thesubappendixc}{\Alph{appendixc}.\arabic{subappendixc}}
\renewcommand{\thesubsubappendixc}
	{\Alph{appendixc}.\arabic{subappendixc}.\arabic{subsubappendixc}}

\renewcommand{\appendix}[1] {\vspace{12pt}
        \refstepcounter{appendixc}
        \setcounter{figure}{0}
        \setcounter{table}{0}
        \setcounter{lemma}{0}
        \setcounter{theorem}{0}
        \setcounter{corollary}{0}
        \setcounter{definition}{0}
        \setcounter{equation}{0}
        \renewcommand{\thefigure}{\Alph{appendixc}.\arabic{figure}}
        \renewcommand{\thetable}{\Alph{appendixc}.\arabic{table}}
        \renewcommand{\theappendixc}{\Alph{appendixc}}
        \renewcommand{\thelemma}{\Alph{appendixc}.\arabic{lemma}}
        \renewcommand{\thetheorem}{\Alph{appendixc}.\arabic{theorem}}
        \renewcommand{\thedefinition}{\Alph{appendixc}.\arabic{definition}}
        \renewcommand{\thecorollary}{\Alph{appendixc}.\arabic{corollary}}
        \renewcommand{\theequation}{\Alph{appendixc}.\arabic{equation}}
        \noindent{\tenbf Appendix \theappendixc #1}\par\vspace{5pt}}
\newcommand{\subappendix}[1] {\vspace{12pt}
        \refstepcounter{subappendixc}
        \noindent{\bf Appendix \thesubappendixc. {\kern1pt \bfit #1}}
	\par\vspace{5pt}}
\newcommand{\subsubappendix}[1] {\vspace{12pt}
        \refstepcounter{subsubappendixc}
        \noindent{\rm Appendix \thesubsubappendixc. {\kern1pt \tenit #1}}
	\par\vspace{5pt}}

\newcommand{\textlineskip}{\baselineskip=13pt}
\newcommand{\smalllineskip}{\baselineskip=10pt}

\def\eightcirc{
\begin{picture}(0,0)
\put(4.4,1.8){\circle{6.5}}
\end{picture}}
\def\eightcopyright{\eightcirc\kern2.7pt\hbox{\eightrm c}}

\def\abstracts#1#2#3{{
	\centering{\begin{minipage}{4.5in}\baselineskip=10pt\footnotesize
	\parindent=0pt #1\par 
	\parindent=15pt #2\par
	\parindent=15pt #3
	\end{minipage}}\par}} 

\renewenvironment{thebibliography}[1]
	{\frenchspacing
	 \ninerm\baselineskip=11pt
	 \begin{list}{\arabic{enumi}.}
	{\usecounter{enumi}\setlength{\parsep}{0pt}
	 \setlength{\leftmargin 12.7pt}{\rightmargin 0pt} 
	 \setlength{\itemsep}{0pt} \settowidth
	{\labelwidth}{#1.}\sloppy}}{\end{list}}

\newcounter{itemlistc}
\newcounter{romanlistc}
\newcounter{alphlistc}
\newcounter{arabiclistc}

\newcommand{\fcaption}[1]{
        \refstepcounter{figure}
        \setbox\@tempboxa = \hbox{\footnotesize Fig.~\thefigure. #1}
        \ifdim \wd\@tempboxa > 5in
           {\begin{center}
        \parbox{5in}{\footnotesize\smalllineskip Fig.~\thefigure. #1}
            \end{center}}
        \else
             {\begin{center}
             {\footnotesize Fig.~\thefigure. #1}
              \end{center}}
        \fi}

\newcommand{\tcaption}[1]{
        \refstepcounter{table}
        \setbox\@tempboxa = \hbox{\footnotesize Table~\thetable. #1}
        \ifdim \wd\@tempboxa > 5in
           {\begin{center}
        \parbox{5in}{\footnotesize\smalllineskip Table~\thetable. #1}
            \end{center}}
        \else
             {\begin{center}
             {\footnotesize Table~\thetable. #1}
              \end{center}}
        \fi}

\def\@citex[#1]#2{\if@filesw\immediate\write\@auxout
	{\string\citation{#2}}\fi
\def\@citea{}\@cite{\@for\@citeb:=#2\do
	{\@citea\def\@citea{,}\@ifundefined
	{b@\@citeb}{{\bf ?}\@warning
	{Citation `\@citeb' on page \thepage \space undefined}}
	{\csname b@\@citeb\endcsname}}}{#1}}

\newif\if@cghi
\def\cite{\@cghitrue\@ifnextchar [{\@tempswatrue
	\@citex}{\@tempswafalse\@citex[]}}
\def\citelow{\@cghifalse\@ifnextchar [{\@tempswatrue
	\@citex}{\@tempswafalse\@citex[]}}
\def\@cite#1#2{{$\null^{#1}$\if@tempswa\typeout
	{IJCGA warning: optional citation argument 
	ignored: `#2'} \fi}}

\def\pmb#1{\setbox0=\hbox{#1}
	\kern-.025em\copy0\kern-\wd0
	\kern.05em\copy0\kern-\wd0
	\kern-.025em\raise.0433em\box0}

\def\fnt#1#2{\footnotetext{\kern-.3em
	{$^{\mbox{\scriptsize #1}}$}{#2}}}

\def\fpage#1{\begingroup
\voffset=.3in
\thispagestyle{empty}\begin{table}[b]\centerline{\footnotesize #1}
	\end{table}\endgroup}

\def\runninghead#1#2{\pagestyle{myheadings}
\markboth{{\protect\footnotesize\it{\quad #1}}\hfill}
{\hfill{\protect\footnotesize\it{#2\quad}}}}
\font\tenrm=cmr10
\font\tenit=cmti10 
\font\tenbf=cmbx10
\font\bfit=cmbxti10 at 10pt
\font\ninerm=cmr9

\font\eightrm=cmr8

\def\qed{\hbox{${\vcenter{\vbox{			
   \hrule height 0.4pt\hbox{\vrule width 0.4pt height 6pt
   \kern5pt\vrule width 0.4pt}\hrule height 0.4pt}}}$}}

\begin{document}

\runninghead{Fundamental Strings as Noncommutative Solitons.} 
{Fundamental Strings as Noncommutative Solitons.}

\normalsize\textlineskip
\thispagestyle{empty}
\setcounter{page}{1}


\begin{flushright}
\begin{small}
hep-th/0010181\\
EFI/00-39 \\
October 2000\\
\end{small}
\end{flushright}
\vspace*{0.17truein}

\fpage{1}

\newcommand {\bea}{\begin{eqnarray}}
\newcommand {\eea}{\end{eqnarray}}
\newcommand {\be}{\begin{equation}}
\newcommand {\ee}{\end{equation}}
\def\IR{{\hbox{{\rm I}\kern-.2em\hbox{\rm R}}}}
\def\IH{{\hbox{{\rm I}\kern-.2em\hbox{\rm H}}}}
\def\IC{{\ \hbox{{\rm I}\kern-.6em\hbox{\bf C}}}}
\def\IZ{{\hbox{{\rm Z}\kern-.4em\hbox{\rm Z}}}}

\centerline{\bf FUNDAMENTAL STRINGS AS NONCOMMUTATIVE SOLITONS
\footnote{Talk presented at STRINGS 2000, july 10-15, 2000, 
University of Michigan at Ann Arbor.}}
\vspace*{0.37truein}
\centerline{\footnotesize FINN LARSEN}
\vspace*{0.015truein}
\centerline{\footnotesize\it Enrico Fermi Institute, 
University of Chicago}
\baselineskip=10pt
\centerline{\footnotesize\it 5640 S. Ellis Ave., Chicago, IL-60637, USA.
\footnote{flarsen@theory.uchicago.edu.}}
\vspace*{0.225truein}


\vspace*{0.21truein}
\abstracts{The interpretation of closed fundamental strings as solitons 
in open string field theory is reviewed. Noncommutativity is introduced 
to facilitate an explicit construction. The tension is computed
exactly and the correct spectrum is recovered at long wave length.}
{}{}


\vspace*{1pt}\textlineskip	

\section{Introduction}	        
\vspace*{-0.5pt}
\noindent
It has recently been realized that strings and branes can be interpreted 
as noncommutative solitons in string field theory. The purpose of this talk 
is to review this development and provide a pedagogical introduction to the 
subject. The discussion is an extended version of  
the actual talk at STRINGS 2000 which in turn was based on\cite{hklm}.

String field theory is designed to mimic many aspects of ordinary 
quantum field theories. It is therefore natural to develop intuition about 
string field theory by applying standard field theory techniques in 
this less familiar context. This motivates the construction of classical 
soliton solutions in string field theory. In this talk the specific goal
is to discuss a class of solitons related to the closed fundamental string. 
It turns out useful to 
introduce noncommutativity in the field theory, as a tool that facilitates 
an explicit construction. The result of the computation will be a soliton 
with tension exactly equal to the tension of the fundamental 
string; and it has the same classical fluctuation spectrum as well. 
These facts provide strong circumstantial evidence that the soliton can
be {\it identified} with the fundamental string. In the course of 
the talk some open questions raised by this interpretation will 
be discussed.
For definiteness the bosonic string theory is considered but virtually 
identical results apply to the superstring case. 

\section{Tachyon Condensation and a First Look at Solitons}
Open string theory by definition supports open string excitations. 
Although it is not usually stressed, the ends of the open strings may
be situated anywhere in spacetime. Recalling that $D$-branes are defined 
as defects where open strings can end, the open string vacuum is 
thus characterized by a space-filling $D$-brane. In this terminology the 
$26$-dimensional perturbative vacuum of open bosonic string theory 
is interpreted as a $D25$-brane. The spectrum of the open strings follows
from standard string theory computations; the result is that the lightest mode
is a tachyon, {\it i.e.} it has negative mass-squared. This 
means the potential of the tachyon field has negative second 
derivative in the perturbative vacuum. As is well-known ({\it e.g.} from
the Higgs phenomenon in the standard model) this kind of tachyon signals an 
instability, there is a true vacuum where the ``tachyon'' field has
acquired an expectation value. The driving force behind the developments 
in the last few years was an 
important insight by Sen\cite{senref}, asserting that the true vacuum 
after tachyon condensation is in fact the standard perturbative 
closed string vacuum, {\it i.e.} the vacuum without the $D25$-brane 
and thus without the open strings. A consequence of this physical picture 
is that the energy liberated by the condensation of the tachyon precisely 
cancels the tension of the $D25$-brane. 

We would like to develop a quantitative description of tachyon condensation
in string theory. For fields with nonvanishing mass, such as the tachyon 
field, a constant field does not satisfy the equations of motion. It is 
therefore clear that tachyon condensation inherently involves off-shell 
properties. 
This is the reason that standard perturbative string theory is 
insufficient to analyze the problem, one must apply string field theory.
The level truncation approximation to the cubic string field theory 
has provided 
convincing evidence that the energy liberated by tachyon condensation
indeed equals that of the $D25$-brane\cite{levelt}. This supports
Sen's identification of the nonperturbative vacuum.

The interest in this talk is the more detailed question of {\it excitations} 
of the non-perturbative closed string vacuum; in particular, those 
adequately described as classical solitons. For 
definiteness the focus will be on the fundamental string excitations. 
Similar considerations apply to other soliton excitations, principally the 
lower-dimensional $D$-branes. The $D$-branes are in fact understood 
more precisely in this set-up; they are the topic of J. Harvey's talk at 
this conference.

The description of the fundamental string in open string theory is 
qualitatively as follows\cite{yi,senqual,hk}. Consider first the situation  
{\it before} tachyon condensation. Then the open strings are
described as some gauge field theory on the world-volume of the 
$D25$-brane. In this framework fundamental
strings appear as electric flux-tubes. {\it After} 
tachyon condensation all open string degress of freedom are removed from 
the spectrum, and in particular the gauge field no longer exists. It is 
therefore not so obvious how to describe the electric flux-tubes after 
tachyon condensation.
This is precisely the problem of interest because, whatever the appropriate 
description of electric flux-tubes after tachyon condensation, these 
are the fundamental strings. The whole process of tachyon condensation 
is reminiscent of the confinement of quarks: the open string degrees of 
freedom cannot propagate in the closed string vacuum but instead manifest
themselves as collective excitations, such as the fundamental strings.

\section{From Noncommutativity to a Quantitative Description}
We would like to turn these comments into a quantitative field theory
description. A suitable starting point for the discussion is
the Born-Infeld Lagrangian\cite{senqual}
\be
S_I = - \int d^{26}x~V(t)~\sqrt{-{\rm det}
[g_{\mu\nu}+2\pi\alpha^\prime F_{\mu\nu}]}~.
\label{eq:bilag}
\ee
One immidiate problem is that the tachyon potential $V(t)$ is unknown (except 
for a few qualitative features discussed in the previous section).
A more pressing concern is that (\ref{eq:bilag}) is justified only for 
{\it constant} fields. In the complete Lagrangean the tachyon has
kinetic terms, and there are numerous higher derivative terms that
generally couple the tachyon and the gauge field. All these terms
are determined {\it in principle} by string field theory. In practice 
they are unfortunately difficult to compute accurately. For example, 
an exact determination of these terms require that infinitely many 
massive fields of the string field are taken into account\footnote{In
the recent works\cite{recbsft} it was noted that the tachyon can be 
decoupled from the massive fields in the BSFT formalism of string field 
theory\cite{bsftrefs}.}. The unknown 
derivative terms are important for fields varying over distances of order 
string scale. They are a serious problem for our purposes because   
the fundamental string solution we seek by definition varies over the string 
scale.

This is the point where noncommutativity turns out useful. 
Recall that a $B$-field can be incorporated in string 
theory by replacing
the standard ``closed string'' metric $g_{\mu\nu}$ and coupling 
constant $g_s$ with the ``open string'' quantities\cite{ncpapers,sw}
\bea
G_{\mu\nu} &=& g_{\mu\nu} - (2\pi\alpha^\prime)^2 (Bg^{-1}B)_{\mu\nu}~,\\
G_s &=& g_s \left( {{\rm det}G\over {\rm det}
(g+2\pi\alpha^\prime B)}\right)^{1\over 2}~.
\eea
More importantly, one must also replace the standard multiplication
of fields with the noncommutative star-product
\be
A\star B = \exp\left({i\over 2}\theta^{\mu\nu}
\partial_\mu\partial_{\nu^\prime} \right)A(x^\mu)B(x^{\prime\nu})~,
\ee
where
\be
\theta^{\mu\nu} = - (2\pi\alpha^\prime)^2 \left( 
{1\over g+2\pi\alpha^\prime B}B
{1\over g-2\pi\alpha^\prime B}\right)^{\mu\nu}~.
\ee
The reason this is useful for the present problem is that the 
noncommutativity parameter $\theta$ provides a new scale in the theory.
The linear extent of the fundamental string solitons we seek are of order
string scale $l_s=\sqrt{\alpha^\prime}$ when measured in the 
{\it closed string metric}; but for large 
$\theta/\alpha^\prime$ this corresponds to a distance $\sqrt{\theta}$ when 
measured in the {\it open string metric}. By way of comparison, the 
complications 
of open string field theory we want to control are of string scale with 
respect to the open string metric; so these are negligible on the scale of 
the soliton, as long as we take the limit $\theta/\alpha^\prime\to\infty$. 
For related discussions see\cite{dmr,witsft}.

Before proceeding with the main line of development it is helpful to discuss 
the limit of large noncommutativity in more detail. Many workers 
(including Seiberg and Witten\cite{sw}) consider $D$-branes in 
background $B$-fields 
and take the low energy decoupling limit
\be
\alpha^\prime \sim \epsilon^{1\over 2} \to 0~~~;~~g_{ij}\sim \epsilon \to 0~,
\label{eq:swlimit}
\ee
for $i,j$ in the noncommutative directions. In this limit string theory 
reduces to noncommutative Yang-Mills theory (with some specific matter 
content). The decoupling limit (\ref{eq:swlimit}) clearly implies large 
noncommutativity
\be
\theta/\alpha^\prime \sim \epsilon^{-{1\over 2}} \to \infty~,
\ee
but it is {\it not} the limit we are considering. We take
$\theta/\alpha^\prime\to\infty$ {\it without} taking the
low energy limit. This is important for our purposes because we want to 
keep string excitations. 
There are two dimensionless parameters in the problem 
$\alpha^\prime E^2$ and $\theta/\alpha^\prime$. The decoupling limit 
(\ref{eq:swlimit}) takes $\theta/\alpha^\prime\to\infty$ with $\theta$,$E$
fixed, and thus
$\alpha^\prime E^2\cdot\theta/\alpha^\prime$ fixed. In contract,
we simply take $\theta/\alpha^\prime\to\infty$ with $\alpha^\prime E^2$
kept fixed. Our limit is that of noncommutative string field theory (NCSFT).
As far as we are aware this limit has not been considered prior to\cite{hklm}.

We now return to the quest for a description of the fundamental string as
a soliton by applying the limit of large noncommutativity to the 
Born-Infeld type Lagrangian (\ref{eq:bilag}). The string solution is going 
to be along some spatial direction, say $x^1$, as well as time $x^0$; 
we take large non-commutativity in all other directions. This 
introduces the open string metric $G^{ij}$, the open string coupling 
constant, and the star product in (\ref{eq:bilag}), yielding
\be
S_I = - {g_s\over G_s}\int d^{26}x~V(t)~\sqrt{-{\rm det}
[G_{\mu\nu}+2\pi\alpha^\prime F_{\mu\nu}]}~.
\label{eq:bilag2}
\ee
The overall factor arises from the replacement $g_s\to G_s$ in the $D$-brane 
potential $V(t)\propto 1/g_s$. The progress at this point is that it is now 
justified to ignore derivatives along all transverse directions. Thus, for 
solutions independent of $x^0,x^1$, we can simply use (\ref{eq:bilag2}) as it 
stands. 

It is convenient for our purposes to consider the conjugate Hamiltonian 
\be
H = - \int d^{25}x~\left[\sqrt{ V(t)^2 + E^2/ (2\pi\alpha^\prime)^2}
+ \lambda \partial_1 E\right] - \lambda p~,
\label{eq:hamil}
\ee
where the electric field $E$ is the canonically conjugate of the gauge
field $A_1$, and $\lambda$ 
is a Lagrange multiplier imposing a restriction to the 
flux sector with quantum number $p$. The equations of motions become
\bea
{V(t)V^\prime(t)\over\sqrt{V(t)^2 + E^2/(2\pi\alpha^\prime)^2}}&=&0
~,
\label{eq:eom1}\\
{E\over\sqrt{V(t)^2 + E^2/(2\pi\alpha^\prime)^2}}+ 
\lambda (2\pi\alpha^\prime)^2 &=& 0~.
\label{eq:eom2}
\eea
These equations are quite simple. In fact, they appear too simple to 
allow nontrivial localized solutions. For example, the equations are  
obviously solved for tachyons with $V^\prime(t)=E=\lambda=0$ 
but this is only possible at the localized extrema of the potential and 
these solutions are therefore constant in spacetime. They are quite different 
from the {\it localized} solitons we seek.

At this point noncommutativity comes to the rescue again\cite{gms}. It is
instructive to consider the equation
\be
\phi * \phi = \phi~.
\label{eq:pp}
\ee
The constant function $\phi=1$ is the only solution,
if $*$ is treated as the ordinary multiplication; but the 
noncommutative *-product involves infinitely many
derivatives so the equation is actually a differential equation 
which may have nontrivial solutions. Indeed, there are many 
solutions\cite{gms}; 
the simplest is the Gaussian
\be
\phi_0 = 2^{12}e^{-r^2/\theta}~,
\label{eq:gauss}
\ee
where $r^2=(x^2)^2+\cdots+(x^{25})^2$.
Solutions to the equation $\phi*\phi =\phi$ are useful because functionals 
act on such functions in a simple way. For any functional $f$ that
can be expanded as a power series we have
\be
f(a\phi) = \sum_{k=0}^\infty c_k a^k \phi^k = f(0) + [f(a)-f(0)]\phi~,
\ee
where $a$ is an ordinary number. As a result of this property the
equations of motion become algebraic for {\it ans\"{a}tze} built
on solutions to $\phi *\phi = \phi$. It is therefore straightforward
to find nontrivial solutions to the equations of motion.

\section{The String Solutions}
Let us consider some simple string solutions obtained this way. 
The simplest possibility is to take
\be
t = t_*\phi_0~,
\label{eq:dstring}
\ee
and other fields vanishing. Here $t_*$ is chosen as the field at the
perturbative extremum of the potential so that\cite{gms} 
\be
V^\prime(t_*\phi_0)=V^\prime(t_*)\phi_0=0~.
\ee
At large distances $\phi_0\to 0$ so $t\to 0$ in the solution 
(\ref{eq:dstring}). In our conventions this corresponds to the
nonperturbative vacuum. The solution (\ref{eq:dstring}) is interpreted 
as a $D$-string and discussed in more detail in\cite{hklm}.

It is simple to verify that 
\be
t = t_* \phi_0 ~;~~ E = p\phi_0~,
\ee
satisfies (\ref{eq:eom1}-\ref{eq:eom2}) as well. Again, the solution is 
essentially the Gaussian (\ref{eq:gauss}); it is therefore fully localized
and asymptotes the closed string vacuum at infinity.
The tension of the string soliton is 
determined from (\ref{eq:hamil}) as
\be
T = {1\over 2\pi\alpha^\prime}\sqrt{{1\over g_s^2}+p^2}~.
\label{eq:p1tension}
\ee
This result suggests that the solitonic string can be identified with
the $(p,1)$ string, {\it i.e.} the bound state of $p$ fundamental
strings and a $D$-string. Repeating the computation starting with
other solutions to $\phi *\phi=\phi$ we find more general
string solutions which can be interpreted as $(p,q)$ strings with 
$q>1$. The main goal is to find a solution with precisely the tension 
of the fundamental string, without any $D$-branes present. This seems
to require a separate consideration. A
candidate string solution with the correct tension 
$T={1\over 2\pi\alpha^\prime}$ is
\be
t=0~~~;~~E=p\phi_0~.
\label{eq:fs}
\ee
Note that in each case discussed above the tension agrees 
{\it exactly}
with the one known from perturbative string theory, even though the 
theory may not be supersymmetric.

The identification of (\ref{eq:fs}) with the fundamental string 
is not entirely unproblematic. One issue is that the fundamental
string tension is independent of $g_s$.
This is puzzling because the action (\ref{eq:bilag})
depends on the coupling only through an overall factor 
$S\propto V(t)\propto 1/g_s$. The key feature that makes this possible is 
that $V(t)=0$ in (\ref{eq:fs}); this invalidates a simple scaling 
argument for the energy. The situation is similar to that of a massless 
particle with $V(t)$ playing the role of mass: the Lagrangian degenerates 
but the Hamiltonian presents no subtleties. Even though the tension 
computation is thus technically sound there is a cause for concern: 
$V(t)\to 0$ suggests that the effective coupling of the problem 
diverges, making quantum corrections important. Our understanding is
that the correct loop counting parameter actually stays well-behaved
so that quantum corrections are under control; however, this point
deserves closer scrutiny.

Let us consider another issue. The solution (\ref{eq:fs}) has the
correct tension and electric flux to be identified with the fundamental 
string. The problem is that many other solutions have the same properties.
Roughly speaking the equations of motion do not constrain the
transverse profile of the solution at all. Specifially, if $\phi_k$ denotes
a complete basis of solutions to (\ref{eq:pp}) then the
configurations
\be
E = \sum_{k=0}^\infty a_k \phi_k ~~~;~~a_k\in\IR~,
\ee
all satisfy the classical equations of motion and could potentially be 
interpreted as the fundamental string. 

General quantum properties of the underlying gauge theory improve the 
situation somewhat by requiring the $a_k$ integral. Even so, countably 
infinite candidate fundamental strings remain. Fortunately this is not 
the end of the story: there is no conserved quantity preventing these 
configurations from mixing quantum mechanically. In fact standard 
maximally supersymmetric D-brane dynamics would provide a unique 
quantum ground state. Unfortunately the situation is more involved here 
and the quantum problem cannot be analyzed precisely, but it may
again have a unique ground state. The quantum problem deserves a better 
understanding.

In this talk we consider only the classical problem. Then the profile 
of the fundamental string is undetermined, it can spread out without 
violating any conservation laws. This is probably the correct result for
infinite flux-tubes. To show flux confinement we need to add electric
sources in bulk and show their flux escapes as a tube, rather than
spreading out like a Coulomb field. It is possible that this can
be understood already classically, as in\cite{kosu}, but we have
not done so.

\section{The Operator Formalism and Finite Noncommutativity}
The presentation of the solutions above can be streamlined and 
the results significantly strengthened by introducing the operator 
formalism. The idea is to exploit the analogy between noncommutative 
geometry and the more familiar noncommutativity in quantum mechanics.
The precise map between the two problems associates an operator
${\hat A}$ to each function $A$ on the noncommutative space, and
maps the noncommutative product $A*B$ to the more familiar operator
product ${\hat A}{\hat B}$ in Hilbert space. In this way problems 
in noncommutative geometry translate into standard exercises in 
quantum mechanics.

Consider as an example the key equation $\phi*\phi=\phi$. In the
operator formalism it reads ${\hat\phi}^2 = {\hat\phi}$ and therefore
its solutions are simply the projection operators in Hilbert space.
It is now clear that the equation has numerous solutions, indeed
infinitely many. The soliton solutions presented in section 4 are
thus essentially projection operators.

The operator formalism makes a huge symmetry manifest. Indeed, 
physics is left invariant under
unitary transformations of operators and states
\be
|\psi\rangle \to U|\psi\rangle~~;~
\langle \psi | \to \langle\psi |U^{\dagger}~~;~
{\cal O}\to U{\cal O}U^{\dagger}~,
\label{untransf}
\ee
where
\be
UU^{\dagger}=U^{\dagger}U=I~.
\label{undef}
\ee
These symmetries form the group $U({\cal H})$. 
This points to a potential embarassment because it shows that any soliton 
solution in the theory has infinitely many ``images'' under $U({\cal H})$.
In many situations we want a unique soliton, to be identifed with its 
counterpart in closed string theory. The crucial observation is that
the $U({\cal H})$ is in fact a {\it gauge symmetry}\cite{lois}. 
The images under $U({\cal H})$ are therefore not interpreted as distinct, 
but as gauge equivalent representations of a single physical state.
In superstring theory the gauge symmetry is further enhanced by an infinite 
discrete group which removes certain tensionless solitons with no reasonable 
physical interpretation\cite{hkl1}.

A key step in the discussion of section 3 was taking the limit
$\theta/\alpha^{\prime}\to\infty$ in order to justify neglecting 
derivative terms. We are now in a position to present an alternative
argument, valid at any $\theta$~\cite{hkl2}. 
The objectionable derivative terms are in fact all 
{\it gauge covariant} derivatives under the $U({\cal H})$ symmetry. We 
can therefore imagine adjusting the gauge fields in the solution precisely 
such that the gauge covariant derivatives vanish identically, removing the 
need for neglecting them. That this is always possible relies on solution
generating transformations of the form (\ref{untransf}) but with
(\ref{undef}) replaced by
\be
UU^{\dagger}=I~~,~U^{\dagger}U = I - P~,
\ee
where $P$ is some projection operator. We can therefore repeat the 
construction of solitons for finite $\theta$~\cite{hkl2}. 
This result is not
surprising: in the closed string vacuum different values of the $B$-field 
are in fact gauge equivalent and it was therefore expected that vacua
with different values of $B$ are related. The argument above shows how this 
works in the open string variables by representing the noncommutative
solitons as ``almost'' gauge equivalent to vacuum, at any $\theta$.

\section{Fluctuations}
The noncommutative Born-Infeld type action (\ref{eq:bilag2}) also 
describes long-wave length fluctuations depending on the commutative 
directions $x^0,x^1$. Allowing for these, the Hamiltonian (\ref{eq:hamil})
is replaced by
\be
H = \int d^{25}x\left[ \sqrt{E^\alpha M_{\alpha\beta}E^\beta
+ V(t)^2 {\rm det}(I+F)}+ A_0 \nabla_\alpha E^\alpha\right]~,
\ee
where
\be
M_{\alpha\beta} = \delta_{\alpha\beta} - F_{\alpha\gamma}F^\gamma_{~\beta}~.
\ee
Using $t=0 \Rightarrow V(t)=0$ (closed string vacuum),
$A_0 = 0$ (gauge condition),
$F_{ij}=0, F_{1i}=A^\prime_i$ (derivatives negligible in NC directions)
we find
\be
H = \int d^{25}x \sqrt{(E^1)^2 ( 1+(\vec{A}^\prime)^2)
+ \vec{E}^2 + (\vec{E}\cdot\vec{A}^\prime)^2 }~,
\ee
where prime denotes the spatial derivative along the string and the 
vector notation refers to the transverse coordinates.
The {\it ansatz} for a fluctuating string is
\bea
\phi &=& \phi_0 (x^i - f^i(x^0,x^1))~,\\
E^1 &=& p\phi_0~,\\ 
\vec{E}&=&\vec{e}\phi_0~,\\ 
\vec{A}^\prime &=& \vec{a}^\prime \phi_0~. 
\label{flucansatz}
\eea
We would like to find the effective action in $D=1+1$ dimensions
controlling the functions $f^i$. The Hamiltonian reduction procedure
accomplishes this, with the result
\be
H = \int dx^1 \sqrt{1 + \vec{\pi}^2 + (\vec{f}^\prime)^2 + 
(\vec{\pi}\cdot\vec{f}^\prime)^2}~.
\ee
It is easy to compute the corresponding Lagrangian. In static gauge 
$X^\mu=(x^0,x^1,f^i)$ one finds
\be
L = - \int d^2 x \sqrt{(\dot{\vec X})^2 (\vec{X}^\prime)^2
- (\dot{\vec{X}}\cdot \vec{X}^\prime)^2 }~.
\label{eq:ng}
\ee
This shows that {\it the effective action of long wave length fluctuations
is the Nambu-Goto action with the correct tension!} The spectrum 
of the soliton is therefore precisely the same as for a fundamental 
string. If we take the action (\ref{eq:ng}) seriously and quantize it we 
find {\it very} light excitations propagating along the string, including 
the graviton and even the closed string tachyon. These would appear 
here as collective excitations in open string field theory, a fascinating 
result. There is a standard objection against this line of reasoning: 
it is only the lightest objects in a theory that can be quantized and 
solitons must therefore usually be treated classically. This
objection fails here because, unlike in many superficially similar 
computations, the fundamental string soliton is indeed the lightest 
excitation of the closed string vacuum, and therefore subject to 
quantization. It is nevertheless unjustified to trust the present 
computation beyond the long-wave length approximation: we are using 
here a crude effective action that only takes into account constant 
fields. It is justified to neglect derivatives in the transverse directions,
because they are noncommutative, but fluctuations in the 
spatial and temporal directions remain. The effective action therefore 
applies only for long wave lengths satisfying 
$\sqrt{\alpha^\prime} F^\prime_{\mu\nu} \ll F_{\mu\nu}$. It is reasonable
to expect that, in a better description, the fluctuations of the 
closed strings can justifiably be interpreted in terms of open string 
variables. The discussion of $\theta$-dependence in section 5 may
be a step in this direction.

\section{Summary}

We conclude with a brief summary. 

\begin{itemize}
\item
The starting point was the process of tachyon condensation in
open string field theory, from the standard perturbative vacuum
to the nonperturbative vacuum, identified with the closed string
vacuum. Our interest was in soliton excitations of the nonperturbative
vacuum.

\item
Noncommutativity was introduced in the system because it generates
a new scale. In the limit of large noncommutativity it is justified
to ignore higher derivatives and use a simple effective field theory 
description of the open string field theory. An argument for ignoring 
the derivatives even at finite $\theta$ was presented in section 5.

\item
An explicit construction of solitons identified with $(p,q)$ strings 
follow. 
Solitons with the exact tension and long wave length fluctuations
of fundamental strings were similarly constructed. These are
identified with the closed fundamental strings expected in the
nonperturbative vacuum.

\item
Several problems remain in the identification of noncommutative
solitons and fundamental strings: the quantum properties of the effective
gauge dynamics need better understanding, confinement must be
elucidated already in the classical description, and the spectrum 
of fluctuations must be computed beyond the long wave length 
approximation. 
\end{itemize}

\nonumsection{Acknowledgements}
\noindent
I am grateful to J. Harvey, P. Kraus, and E. Martinec for collaboration 
yielding the results reported here, among others. 
This work was supported by DOE grant DE-FG0290ER-40560 and by a Robert 
R. McCormick fellowship. 

\nonumsection{References}

\end{document}